\documentclass[aps, prd, twocolumn, lengthcheck, superscriptaddress, showpacs, letterpaper, nofootinbib]{revtex4-1}
\usepackage{amsfonts}
\usepackage{amsmath}
\usepackage{amssymb}
\usepackage{xcolor}
\usepackage{graphicx}
\usepackage{longtable}
\usepackage{slashed}
\usepackage{hyperref}
\usepackage{amsthm}
\usepackage{mathrsfs}
\usepackage{color}

\def\be{\begin{eqnarray}}\def\ee{\end{eqnarray}}

\allowdisplaybreaks

\begin{document}
\title{Chiral effective Lagrangian for heavy-light mesons from QCD}

\author{Qing-Sen Chen}
\affiliation{Center for Theoretical Physics, College of Physics, Jilin University,
Changchun 130012, China}

\author{Hui-Feng Fu}
\email{huifengfu@jlu.edu.cn}
\affiliation{Center for Theoretical Physics, College of Physics, Jilin University,
Changchun 130012, China}

\author{Yong-Liang Ma}
\email{yongliangma@jlu.edu.cn}
\affiliation{Center for Theoretical Physics, College of Physics, Jilin University,
Changchun 130012, China}
\affiliation{School of Fundamental Physics and Mathematical Sciences,
Hangzhou Institute for Advanced Study, UCAS, Hangzhou, 310024, China}
\affiliation{International Centre for Theoretical Physics Asia-Pacific (ICTP-AP) (Beijing/Hangzhou), UCAS, Beijing 100190, China}

\author{Qing Wang}
\email{wangq@mail.tsinghua.edu.cn}
\affiliation{Department of Physics, Tsinghua University, Beijing 100084, China 
}
\affiliation{Center for High Energy Physics, Tsinghua University, Beijing 100084, China 
}


\begin{abstract}
We derive the chiral effective Lagrangian for heavy-light mesons in the heavy quark limit from QCD under proper approximations. The low energy constants in the effective Lagrangian are expressed in terms of the light quark self-energy. With typical forms of the running coupling constant of QCD and the quark self-energy obtained from Dyson-Schwinger equations as well as lattice QCD, we estimate the low energy constants in the model and the strong decay widths. A comparison with data and some discussions of the numerical results are presented.
\end{abstract}
%

 \maketitle

\section{Introduction}

It is widely accepted that the dynamics of strong interaction at a large distance can be well captured by effective theories which are controlled by certain symmetries and symmetry breaking, for example chiral symmetry breaking, of the fundamental theory of strong interaction---QCD (see, e.g., Ref.~\cite{Hatsuda:1994pi}). Little progress has been made to establish a direct relation between QCD and an effective theory in an analytic way although such a relation is very significant since, to our opinion, almost all the puzzles and problems in hadron dynamics in both matter free space and in-medium can be attributed to our poor understanding of the nonperturbative QCD.

In Ref.~\cite{Wang:1999cp}, the authors derived a relation between QCD and chiral perturbation theory (ChPT) including the Nambu-Goldstone bosons, pions, only~\cite{Weinberg:1978kz,Gasser:1983yg,Gasser:1984gg}. The derivation is based on the standard generating functional of QCD with external bilinear light-quark field sources in the path integral formalism so that, the coefficients in the chiral Lagrangian can be derived from QCD Green functions. Thanks to the developments of Dyson-Schwinger equation (DSE) and lattice QCD methods, although an exact calculation of the nonperturbative QCD Green functions is not feasible so far, this derivation provides a possibility to numerically calculate the low energy constants (LECs) of ChPT from QCD---the fundamental theory of strong interaction~\cite{Yang:2002hea,Jiang:2009uf,Jiang:2010wa,Jiang:2012ir,Jiang:2015dba}. The extension of this approach to include other light mesons are also investigated in the literature~\cite{Wang:2000mu,Wang:2000mg,Wang:2000mu,Ren:2017bhd}.

The purpose of this work is to establish a relation between fundamental QCD and chiral effective theory for heavy-light mesons~\cite{Wise:1992hn} in the widely accepted heavy quark limit, although the extension to include the $1/M$ correction---with $M$ being the heavy quark mass---is straightforward. Since in the heavy-light meson, there is a heavy quark and a light quark, its dynamics is controlled by both the heavy quark symmetry~\cite{Isgur:1991wq} and chiral symmetry (see Ref.~\cite{Manohar:2000dt} for a systematic discussion). This system is a good environment for studying the mechanism of chiral symmetry breaking since it contains only one light quark.

In addition to the lowest-lying heavy quark doublet $H$ with quantum numbers $J^P=(0^-,1^-)$, we introduce the first orbital excitation states, the heavy quark $G$ doublet with quantum numbers $J^P=(0^+,1^+)$ (both $H$ and $G$ doublets will be specified latter) to our system. Since in these $H$ and $G$ doublets, the light quark clouds have quantum numbers $j^P=(1/2)^+$ and $j^P=(1/2)^-$, respectively, they are regarded as chiral partners to each other, and their mass splitting arises from the chiral symmetry breaking due to the quark condensation in the chiral doublet model~\cite{Nowak:1992um,Bardeen:1993ae}. We explore the quark condensate dependence of the mass splitting between the $H$ and $G$ doublets in the present approach from fundamental QCD.

We express the LECs in the effective theory, for example the heavy-light meson mass, the coupling constant between the heavy-light meson and pion, in terms of the light quark self-energy (with the heavy quark mass rotated away), therefore the LECs become calculable from fundamental QCD. By using the light-quark self-energy calculated from truncated DSE and lattice QCD, we calculate these LECs as functions of quark condensate. Both the mass spectrum of the heavy-light mesons and the one-pion transition obtained in our calculation are comparable to the empirical values. In addition, we find that, with the decreasing of the quark condensate, the masses of both the $H$ and $G$ doublets decrease. Although this observation agrees with Ref.~\cite{Harada:2016uca} with a specific choice of the parameters in the model, it
disagrees with what was found in Ref.~\cite{Suenaga:2014sga} that although the $G$ doublet mass decreases with the decreasing of the quark condensate, the $H$ doublet mass increases. So far, we cannot give a clear comment on this discrepancy between effective model calculation and the present QCD approach. What is interesting in the present calculation is that the mass splitting between $H$ and $G$ doublets decreases with the decreasing of the quark condensate. This tells us that the mass splitting between chiral partners arises from the quark condensate, at least partially if not all. A more possibly interesting point can be drawn from the present work is that, once the quark condensate is influenced by the environment such as the medium density and temperature, the LECs in the effective Lagrangian are also affected. In the literature, such modification is referred to as intrinsic density dependence in the dense system as opposite to the density dependence induced by nuclear correlations~(see, e.g., Ref.~\cite{Ma:2019ery} for a recent comprehensive discussion).

This paper is organized as follows. In Sec.~\ref{sec:HeavyEFT}, to set up the framework, we write down the chiral effective theory of the heavy-light mesons that will be derived in the present work. In Sec.~\ref{sec:EFTQCD} we derive the heavy-light meson Lagrangian from QCD and express the low energy constants in the effective theory in terms of the light quark self-energy. The numerical results calculated by using the quark self-energy obtained from a typical DSE and lattice QCD are given in Sec.~\ref{sec:Num}. Section~\ref{sec:dis} is devoted to our discussion and perspective. We present some details of the derivation in Appendix~\ref{sec:AppA} and the expressions of the LECs with the contribution from the renormalization factor of quark wave function in Appendix~\ref{sec:AppB}.

\section{chiral effective theory of heavy-light mesons}

\label{sec:HeavyEFT}

To set up our framework, we write down the chiral effective theory of heavy-light mesons that will be studied in this work. We introduce the charmed heavy-light meson doublets $H$ and $G$ with quantum numbers $J^P=(0^-, 1^-)$ and $J^P = (0^+, 1^+)$, respectively. In terms of the physical states and the notation of PDG~\cite{ZylaPDG}, they are expressed as
\begin{eqnarray}
H & = & \frac{1 + v\hspace{-0.15cm}\slash}{2}\left(D^{\ast;\mu}\gamma_\mu + i D \gamma_5\right),\nonumber\\
G & = & \frac{1 + v\hspace{-0.15cm}\slash}{2}\left(D^{\prime;\mu}_1\gamma_\mu\gamma_5 + D^\ast_0\right).
\label{eq:HeavyMeson}
\end{eqnarray}
In the following, without specification, we will focus on two light flavors although the extension to three flavors is straightforward.

Under chiral transformation, the $H$ doublet transforms under the unbroken $SU(2)_V$ subgroup of chiral symmetry as an antidoublet
\begin{eqnarray}
H_a & \to &  H_bV_{ba}^\dagger,
\end{eqnarray}
with $a$ and $b$ being the light flavor indices. The same transformation holds for the $G$ doublet.

Then the simplest effective Lagrangian describing the interaction between pion and the heavy-light mesons can be written as~\cite{Manohar:2000dt,Nowak:2004jg}
\begin{eqnarray}
{\cal L} & = & {\cal L}_H + {\cal L}_G + {\cal L}_{HG},
\label{eq:LHHChPT}
\end{eqnarray}
where
\begin{eqnarray}
{\cal L}_H & = & {}-i{\rm Tr} \left(\bar{H} v\cdot D H\right) - g_H {\rm Tr} \left(H \gamma^\mu\gamma_5 A_\mu  \bar{H}\right) \nonumber\\
& &{} + m_H{\rm Tr} \left(\bar{H} H\right),\nonumber\\
{\cal L}_G & = &{} - i {\rm Tr} \left(\bar{G} v\cdot D G\right) + g_G {\rm Tr} \left(G \gamma^\mu\gamma_5 A_\mu  \bar{G} \right) \nonumber\\
& &{} + m_G{\rm Tr} \left(\bar{G} G\right), \nonumber\\
{\cal L}_{HG} & = & g_{HG}{\rm Tr} \left(H \gamma^\mu\gamma_5 A_\mu  \bar{G}\right) + H.c.,
\end{eqnarray}
where we have decomposed the chiral field $U(x) = \exp(i\pi(x)/f_\pi)$ as $U = \Omega^2$ and $A_\mu = \frac{i}{2}\left(\Omega^\dagger \partial_\mu \Omega - \Omega \partial_\mu \Omega^\dagger\right)$ and $D_\mu = \partial_\mu - i \Gamma_\mu$ with $\Gamma_\mu = \frac{i}{2}\left(\Omega^\dagger \partial_\mu \Omega + \Omega \partial_\mu \Omega^\dagger\right)$.

From PDG~\cite{ZylaPDG}, one can obtain the spin-averaged masses of $H$ and $G$ doublets as
\begin{eqnarray}\label{eq:EXmass}
m_H & \simeq & 1970~{\rm MeV}, \;\;\; m_G \simeq 2400~{\rm MeV},
\end{eqnarray}
which yields the mass splitting
\begin{eqnarray}
\Delta m & = & m_G - m_H = 430~{\rm MeV}.
\label{eq:splitting}
\end{eqnarray}
It is believed that the value of the mass splitting between chiral partners arises from the chiral symmetry breaking.

In the chiral doublet structure, the mass splitting $\Delta m$ is attributed to the chiral symmetry breaking, i.e., the vacuum expectation value of the sigma field in the linear sigma model~\cite{Nowak:1992um,Bardeen:1993ae}. Therefore, the magnitude of this mass splitting measures the magnitude of the chiral symmetry breaking, i.e., the larger $\Delta m$, the stronger chiral symmetry breaking.

The Lagrangian~\eqref{eq:LHHChPT} is written down with respect to the chiral symmetry and heavy quark symmetry of QCD. Only the terms with the minimal number of derivatives are included. The parameters $g_H, g_G, g_{HG}, m_H$, and $m_G$ are free ones at the level of effective theory since they cannot be controlled by symmetry argument which the effective theory relies on. We next evaluate these LECs from fundamental QCD.

\section{chiral effective lagrangian of heavy-light mesons from QCD}

\label{sec:EFTQCD}

We derive the heavy-light meson effective theory~\eqref{eq:LHHChPT} from the generating functional of QCD (with an external source $J(x)$ introduced for the composite light quark fields)
\begin{eqnarray}
Z[J] & = & \int \mathcal{D}q \mathcal{D} \bar q \mathcal{D}Q \mathcal{D} \bar Q \mathcal{D} G_{\mu}\varDelta_F(G_{\mu})\nonumber\\
& &\times \exp\left\{i\int d^4 x\left[ {\mathcal{L}_{\mathrm{QCD}}(q,\bar q,Q, \bar Q ,G_{\mu})+\bar q Jq}\right]\right\},\nonumber\\
\label{generating0}
\end{eqnarray}
where $q(x), Q(x)$ and $G_\mu(x)$ are the light-, heavy-quark fields and gluon fields, respectively. By integrating in the chiral field $U(x)$ and the heavy-light meson fields, and integrating out gluon fields and quark fields, we obtain the effective action for the chiral effective theory as
\begin{eqnarray}
\label{action2}
S[U,\Pi_2,\bar{\Pi}_2] & = &
{} -i N_c \mathrm{Tr}^\prime \ln\left\{\left(i\slashed{\partial}-\bar{\Sigma}\right)I_1 +J'_\Omega \right.\nonumber\\
& &\left. {} \qquad\qquad\qquad + \left(i\slashed{\partial}+M\slashed v-M\right)I_4\right.\nonumber\\
& &\left. {} \qquad\qquad\qquad  -\Pi_2-\bar{\Pi}_2 \right\},
\end{eqnarray}
where $\Pi_2$ and $\bar{\Pi}_2$ are the heavy-light meson field and its conjugate, respectively. $\bar{\Sigma}$ is the self-energy of the light quark propagator. $I_1=\mathrm{diag}(1,1,0)$ and $I_4=\mathrm{diag}(0,0,1)$ are matrices in the flavor space. $J_\Omega$ is the chiral rotated external source and $J'_\Omega$ is its extension to the whole flavor space (including the heavy flavor). $\mathrm{Tr}^\prime$ represents a functional trace over the flavor space, spinor space, and coordinate space. To obtain the effective action,
we have taken advantage of both the chiral symmetry and heavy quark symmetry, and made a few approximations. The details of the derivation are given in Appendix \ref{sec:AppA}.

The dynamical content of action (\ref{action2}) could be understood easily once we rewrite $e^{iS}$ with fermionic fields reintroduced:
\begin{eqnarray}
\label{action3}
 e^{iS} & \sim & \int \mathcal{D} q\mathcal{D} \bar q\mathcal{D}  Q \mathcal{D}\bar  Q \exp\biggl\{i\int d^4x \biggl[ \bar  q (i\slashed{\partial}-\bar{\Sigma}+J_\Omega) q\notag\\
 &&+ \bar  Q (i\slashed{\partial}+M\slashed v-M)Q-\bar{Q}\Pi q-\bar{q}\bar{\Pi}Q\biggr]\biggr\}.
\end{eqnarray}
From this expression one can easily conclude that the action (\ref{action2}) is just the summation of quark loops with all possible insertions of $J_\Omega$ and heavy-light meson fields $\Pi$, $\bar{\Pi}$. In addition, one can see that the contribution of gluon fields are included in the self-energy $\bar{\Sigma}$.

We are only interested in the $(0^-, 1^-)$ and $(0^+, 1^+)$ states, so that the two nonzero elements in $\Pi_2$ take the form
\begin{eqnarray}
\Pi_{2}(x) & = & H^q(x)+G^q(x), 
\end{eqnarray}
where $q=1, 2$ denotes light flavor indices and $H$ and $G$ fields are given by Eq.~\eqref{eq:HeavyMeson}.

Expanding the action $S$~\eqref{action2} with respect to $U$, $\Pi_2$ and $\bar{\Pi}_2$ generates the chiral effective Lagrangian. Since the $U$ field is attached to the rotated external source, we actually take derivatives on the action $S$ with respect to $J_\Omega$. So that we can formally write
\begin{eqnarray}
S=S_0+S_1+S_2+S_3+\cdots,
\label{eq:expansionS}
\end{eqnarray}
where the subscripts $0,1,2,3,\cdots$ denote the order in the expansion in terms of $J_\Omega$, $\Pi_2$ and $\bar{\Pi}_2$. In the expansion \eqref{eq:expansionS}, the $S_0$ term, a constant of the action $S$, has no physical significance and, the first order term $S_1$ vanishes due to the saddle point equation. Therefore, the leading order contribution comes from $S_2$. Due to the symmetry arguments, only the following terms survive:
\begin{eqnarray}
S_2=\frac{1}{2!} \frac{\delta^2 S}{\delta J_\Omega \delta J_\Omega} J_\Omega J_\Omega+\frac{\delta^2 S}{\delta \bar{\Pi}_2 \delta \Pi_2} \Pi_2\bar{\Pi}_2.
\end{eqnarray}
The first term generates the nonlinear sigma model of the Nambu-Goldstone bosons which has been extensively discussed in Ref.~\cite{Wang:1999cp}. We will not go to the details here. The second term of $S_2$ generates the model of the heavy-light mesons which is interested in the present work.

\begin{figure*}[!t]
	\centering	
\includegraphics[height=3.5cm]{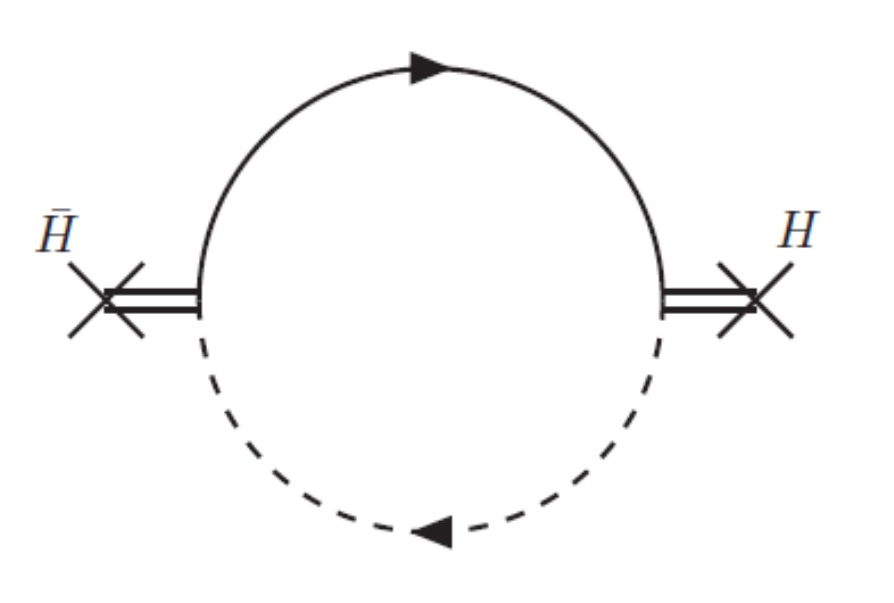}
\includegraphics[height=5.0cm]{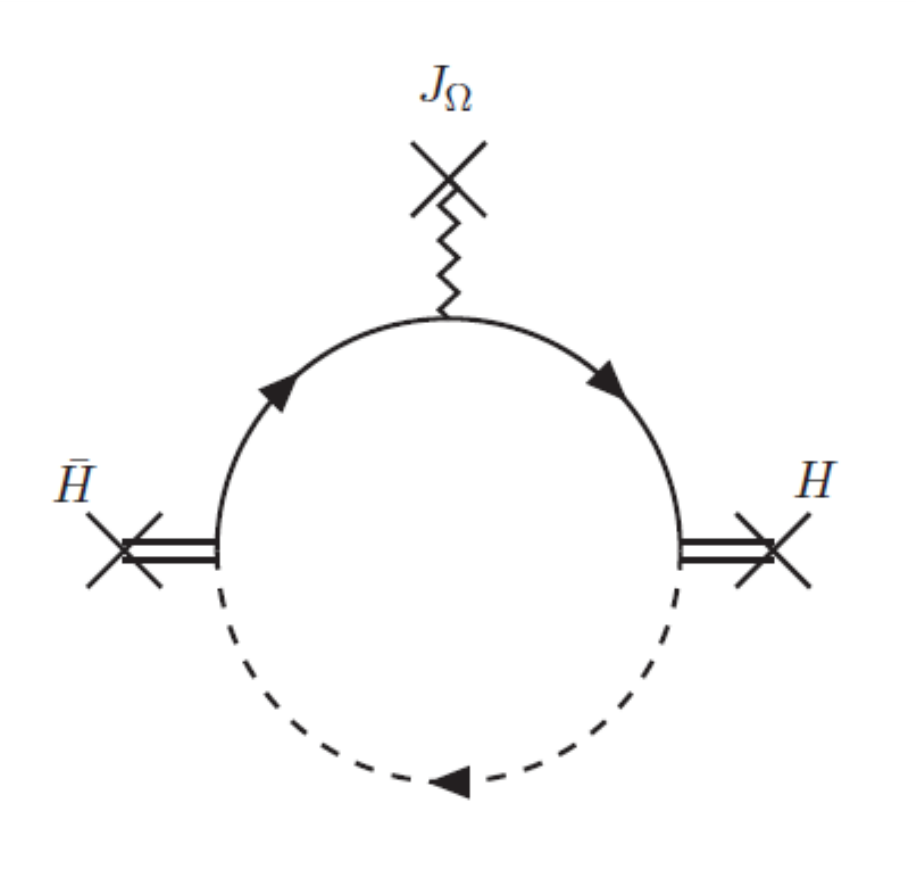}
\caption{Left panel: diagrammatic expression for Eq. (\ref{S22}), i.e., one heavy-light quark loop with one $H$ and one $\bar{H}$ insertions; Right panel: diagrammatic expression for Eq. (\ref{S3}), i.e., one heavy-light quark loop with one $H$, one $\bar{H}$ and one $J_\Omega$ insertions. The solid arrow line represents the full light-quark propagator. The dashed arrow line represents the heavy-quark propagator (in the heavy quark limit). The doubled line attached with a cross represents the $H$/$\bar{H}$ insertion. The zigzag line attached with a cross represents the $J_\Omega$ insertion.}	
\label{oneloop}
\end{figure*}

We denote the second term in $S_2$ as $S_{22}$, then for the fields $H$ and $\bar{H}$, we have
\begin{widetext}
\begin{eqnarray}
S_{22} & = & iN_c\left[\left(\left(i\slashed{\partial} -\bar{\Sigma}\right)I_1+\left(i\slashed{\partial}+M\slashed v-M\right) I_4\right)^{-1}\right]^{b_2 a_1}_{ \xi_2 \eta_1}(x_2,x_1)\bar H^{a_1b_1}_{\eta_1 \xi_1}(x_1) \notag \\
& & \qquad \times \left[\left(\left(i\slashed{\partial} -\bar{\Sigma}\right)I_1+\left(i\slashed{\partial}+M\slashed v-M\right) I_4\right)^{-1}\right]^{b_1 a_2}_{ \xi_1 \eta_2 }(x_1,x_2)  H^{a_2 b_2}_{\eta_2 \xi_2}(x_2), \notag \\
& = & iN_c\int d^4x_1 d^4 x_2 \mathrm{tr}_{lf}\left[\left(i\slashed{\partial} -\bar{\Sigma}\right)^{-1}\delta(x_2-x_1)\bar{H}(x_1)(i v\cdot \partial)^{-1}\delta(x_1-x_2) H(x_2)\right],\label{S22}
\end{eqnarray}
where we have used the identity $\bar H i \slashed \partial H=\bar H i v \cdot \partial H$ and $ \slashed v H=H$. Equation (\ref{S22}) is expressed diagrammatically in Fig. \ref{oneloop} (left panel), which is simply the heavy-light quark loop with one $H$ and one $\bar{H}$ inserted properly. The mass and the kinetic terms of the heavy-light meson are dynamically generated by this quark loop. In order to keep the invariance of the vector part of the chiral transformation, we follow Ref.~\cite{Yang:2002hea} and take the form of the self-energy of the light quark as $\bar{\Sigma}(x-y)=\Sigma(\nabla^2)\delta(x-y)$, where $\Sigma$ is the self-energy function in the momentum space and $\nabla\equiv\partial-iV_\Omega$ is the covariant derivative. This form retains the correct chiral transformation properties in the theory. By taking the derivative expansion, we obtain (up to the first order)
\begin{eqnarray}
S_{22} & = & i N_c\int d^4x \int\frac{d^4  p}{(2\pi)^4} \left[-\frac{1}{ p^2-\Sigma^2} +\frac{\Sigma}{( p^2-\Sigma^2)v\cdot  p}\right] \mathrm{Tr}\left[\bar H(x)H(x)\right]\notag\\
& & {} ~~~~+iN_c\int d^4x \int\frac{d^4  p}{(2\pi)^4} \left[-\frac{1}{( p^2-\Sigma^2) v\cdot  p}- \frac{2\Sigma}{( p^2-\Sigma^2)^2}-\frac{2\Sigma'( p^2+\Sigma^2)}{( p^2-\Sigma^2)^2}\right]\mathrm{Tr}\left[ H(x)iv\cdot\partial \bar H(x)\right]\notag\\
& &{} ~~~~+iN_c\int d^4x \int\frac{d^4  p}{(2\pi)^4} \left[-\frac{2\Sigma'( p^2+\Sigma^2)}{( p^2-\Sigma^2)^2}\right]\mathrm{Tr}\left[ H(x)v\cdot V_\Omega \bar H(x)\right],
\label{eq:S22}
\end{eqnarray}
\end{widetext}
where we have used the relation $H \slashed v =-H$. From Eq.~\eqref{eq:S22} one can easily see that $S_{22}$ includes the mass term, the kinetic term and a part of the interaction term of the $H$ fields.

We next consider the part of the $S_{3}$ which generates the interaction between the heavy-light meson field $H$ and the light Goldstone boson field up to the leading order of the chiral counting. This can be obtained as
\begin{widetext}
\begin{eqnarray}\label{S3}
S_{3} & = & \frac{\delta^3 S}{\delta J_\Omega  \delta \bar H \delta  H } J_\Omega H\bar{H},\notag\\
& = &{} -iN_c\int d^4x_1d^4x_2d^4x_3\mathrm{Tr}\left[(i\slashed{\partial} -\Sigma)^{-1}\delta (x_2-x_3)J(x_3) (i\slashed{\partial} -\Sigma)^{-1} \delta(x_3-x_1) \bar H(x_1)  (iv\cdot\partial)^{-1} \delta(x_1-x_2)  H (x_2)\right].\nonumber\\
\end{eqnarray}
Recalling that $S_\Omega$ and $P_\Omega$ are of $\mathcal{O}(p^2)$ while $A_\Omega^\mu$ and $V_\Omega^\mu$ are of $\mathcal{O}(p)$ in the chiral counting, we neglect the $S_\Omega$ and $P_\Omega$ terms. The diagram representation of Eq. (\ref{S3}) is shown in Fig. \ref{oneloop} (right panel). Taking the derivative expansion, we obtain
\begin{eqnarray}
S_{3} & = &{} -iN_c\int d^4x \int\frac{d^4 p}{(2\pi)^4}\left[\frac{\Sigma^2+\frac{1}{3}p^2}{(p^2-\Sigma^2) v\cdot p}  - \frac{2\Sigma}{(p^2-\Sigma^2)^2 } \right] \mathrm{Tr}\left[ H(x) \gamma_{\mu}\gamma_5 A^{\mu}_{\Omega}(x) \bar H(x)\right]\notag\\
& &{} -iN_c\int d^4x \int\frac{d^4 p}{(2\pi)^4} \left[\frac{1}{(p^2-\Sigma^2) v\cdot p} +\frac{2\Sigma }{(p^2-\Sigma^2)^2} \right]  \mathrm{Tr}\left[ H(x) v_{\mu} V^{\mu}_{\Omega}(x) \bar H(x)\right].
\label{eq:SJHH}
\end{eqnarray}
\end{widetext}

Summing up Eqs.~\eqref{eq:S22} and \eqref{eq:SJHH}, one can obtain the expressions of the constants $m_H$ and $g_H$ as
\begin{eqnarray}
m_H & = & \frac{iN_c}{Z_H}  \int\frac{d^4 p}{(2\pi)^4} \left[{}\frac{1}{p^2-\Sigma^2} - \frac{\Sigma}{(p^2-\Sigma^2)v\cdot p}\right],  \nonumber\\
g_{H} & = &{} -\frac{iN_c}{Z_H}\int\frac{d^4 p}{(2\pi)^4}\left[\frac{\Sigma^2+\frac{1}{3}p^2}{(p^2-\Sigma^2)^2 v\cdot p}  - \frac{2\Sigma}{(p^2-\Sigma^2)^2 } \right],
\nonumber\\
\label{eq:mHgH}
\end{eqnarray}
with $Z_H$ being the wave function renormalization factor
\begin{eqnarray}
Z_H & = & iN_c\int\frac{d^4 p}{(2\pi)^4} \biggr[{} -\frac{1}{(p^2-\Sigma^2) v\cdot p} \nonumber\\
& &{}\qquad\qquad\qquad - \frac{2\Sigma}{(p^2-\Sigma^2)^2} -\frac{2\Sigma'( p^2+\Sigma^2)}{( p^2-\Sigma^2)^2}\biggr].\nonumber
\end{eqnarray}
It is interesting to note that the summation of the $S_{22}$ and $S_3$ terms yields the same coefficients for the $\mathrm{Tr}\left[ H(x)iv\cdot\partial \bar H(x)\right]$ term and $\mathrm{Tr}\left[ H(x)v\cdot V_\Omega \bar H(x)\right]$ term. This means that the vector part of the chiral symmetry is reserved in our approach, in agreement with the pattern of the chiral symmetry breaking in QCD.

By using the same argument, we obtain the LECs of the heavy-light mesons in the positive parity sector. The expressions are
\begin{eqnarray}
m_G & = & \frac{iN_c}{Z_G}\int\frac{d^4 p}{(2\pi)^4} \left[\frac{1}{p^2-\Sigma^2} +\frac{\Sigma}{(p^2-\Sigma^2)v\cdot p}\right], \nonumber\\
g_{G} & = &{} -\frac{iN_c}{Z_G}\int\frac{d^4 p}{(2\pi)^4}\left[\frac{\Sigma^2+\frac{1}{3}p^2}{(p^2-\Sigma^2)^2 v\cdot p}  + \frac{2\Sigma}{(p^2-\Sigma^2)^2 } \right],
\nonumber\\
\label{eq:mGgG}
\end{eqnarray}
with $Z_G$ being the renormalization factor of the $G$ field
\begin{eqnarray}
Z_G & = & iN_c\int\frac{d^4 p}{(2\pi)^4} \biggr[{} -\frac{1}{(p^2-\Sigma^2) v\cdot p} \nonumber\\
& &{} \qquad\qquad\qquad + \frac{2\Sigma}{(p^2-\Sigma^2)^2}+\frac{2\Sigma'( p^2+\Sigma^2)}{( p^2-\Sigma^2)^2} \biggr].\nonumber
\end{eqnarray}
And, we obtain the coupling constant between the parity partner fields $H$ and $G$ as
\begin{eqnarray}
g_{H G} & = & {} - i \frac{N_c}{\sqrt{Z_H Z_G}}\int\frac{d^4 p}{(2\pi)^4}\biggl[\frac{\Sigma^2+p^2}{(p^2-\Sigma^2)^2 v\cdot p} \biggr].
\label{eq:gHG}
\end{eqnarray}
In above expressions, $\Sigma(-p^2)$ stands for self-energy of light quarks to be calculated from QCD. This is an improvement compared to Ref.~\cite{Nowak:1992um}. In addition, our procedure of deriving the chiral Lagrangian allows a systematic improvement in the calculation.

\begin{figure*}[!t]
	\centering	
\includegraphics[height=3.5cm]{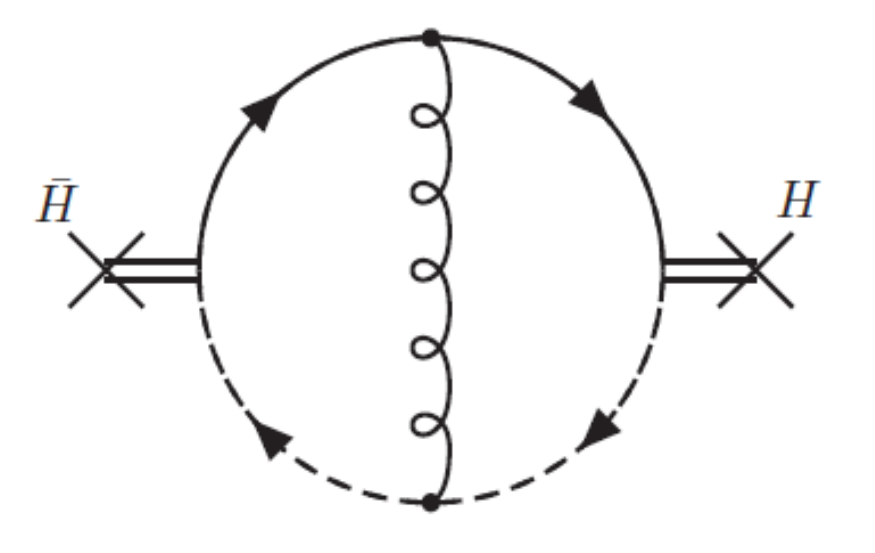}
\includegraphics[height=3.5cm]{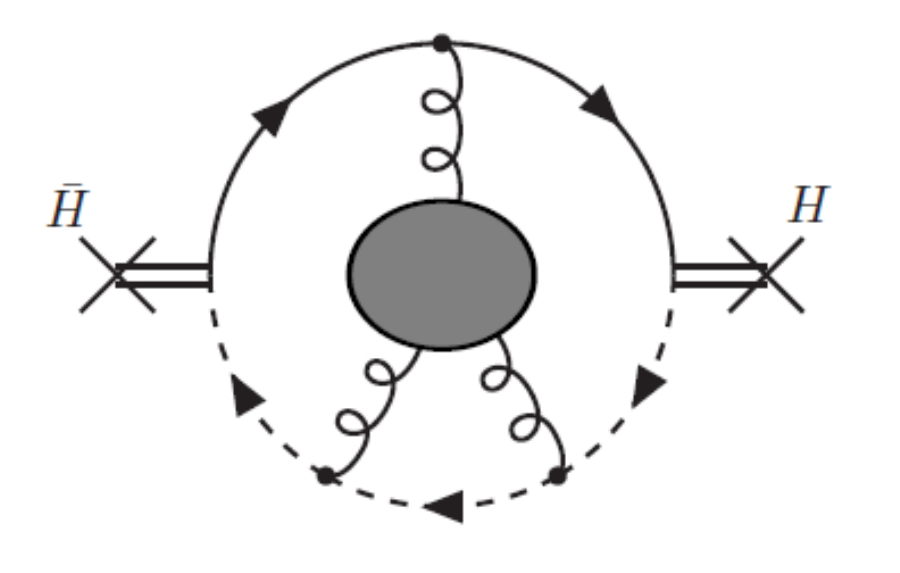}
\includegraphics[height=3.5cm]{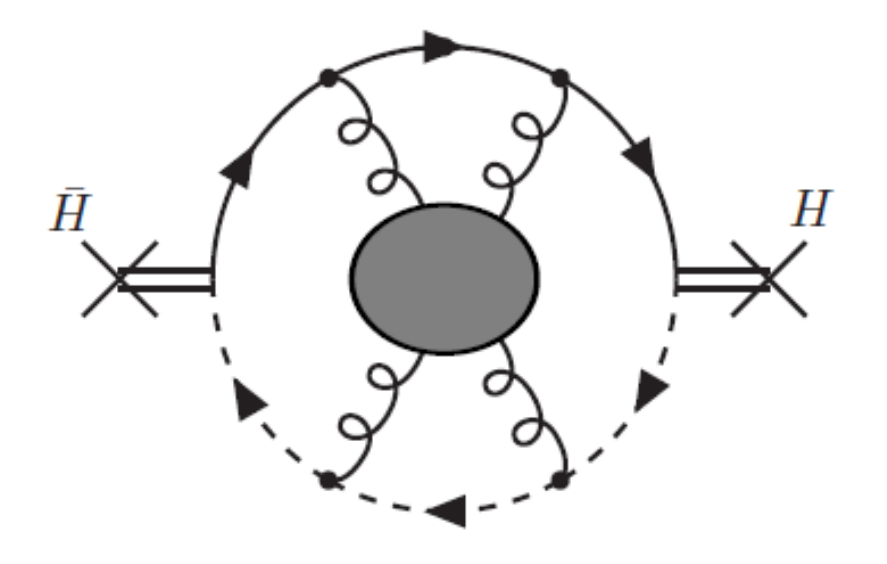}
\caption{Examples of higher loop diagrams contributing to $\bar{H} H$ term in the effective Lagrangian. The gluon propagators are the (quenched) full ones. The shaded ovals with gluon legs represent (quenched) full connected gluon Green functions.}	
\label{higherloops}
\end{figure*}

The LECs discussed so far only receive contributions from one loop diagrams, because we have been using a simplified effective action $S$ (\ref{action2}). The contributions to these LECs from the other effects are too complicated to be quantitatively analyzed, so we only make some qualitative discussions here: In deriving $S$ (\ref{action2}), we took three approximations, namely, the chiral limit, the heavy quark limit and the large $N_c$ limit. So, our results suffer from corrections of $\mathcal{O}(m_{u/d})$, of $\mathcal{O}(1/m_Q)$ and of $\mathcal{O}(1/N_c)$ order. In addition, the non-``$\mathrm{Tr}\ln$'' terms existing in the full large $N_c$ action [see Eq. (\ref{action1})] have been omitted in action $S$ (\ref{action2}) in the sense of dynamical pertubation which has been found workable for the calculation of the pion decay constant~\cite{Pagels:1979hd}. The dynamical content of these non-$\mathrm{Tr}\ln$ terms is relatively simple for the $\bar{H} H$ term (but not for all the terms) in the effective Lagrangian, because only the $\bar{G}(\Phi_c)$ term in Eq. (\ref{action1}) contributes. One can easily see that the $\bar{G}(\Phi_c)$ term is actually the summation of the vacuum diagrams which are composed of one connected full gluon Green's function (pure Yang-Mills) and a number of quark loops with quark propagating on the background of meson fields [$\Pi(x)$ and $U(x)$]. This term gives rise to the contribution from higher-loop diagrams. For illustration, we show a few of those diagrams contributing to the $\bar{H} H$ term in the effective Lagrangian in Fig.~\ref{higherloops}.~\footnote{The $\bar{G}(\Phi_c)$ term given in Eq. (\ref{action1}) only contains $(1/N_c)^0$ order contributions due to its definition. Higher order $\mathcal{O}(1/N_c)$ contributions of this term should arise when we go beyond the large $N_c$ limit. So the diagrams given in Fig. \ref{higherloops} give rise to higher loop corrections as well as $1/N_c$ corrections.} We shall not calculate these contributions in the present work, but stick to the leading order contributions given by action $S$ (\ref{action2}).

\section{Numerical results }

\label{sec:Num}

We have obtained the chiral effective Lagrangian for the heavy-light mesons and the integral forms of the LECs in terms of the dynamical quark mass $\Sigma(-p^2)$. To have a quantitative idea, we make some numerical calculations in this section. For this purpose, we should know the light quark self-energy $\Sigma(-p^2)$, of which only the information from DSEs and lattice QCD is available at this moment.

First, we consider the information from the DSE method. Here, we take the following differential form of DSE for the quark self-energy~\cite{Yang:2002hea}~\footnote{There might be a tiny difference of the numerical values given below if we take other forms. But it does not affect the main conclusion of this work. }
\begin{eqnarray}
& &\left(\frac{\alpha_s(x)}{x}\right)'\Sigma(x)''-\left(\frac{\alpha_s(x)}{x}\right)'' \Sigma(x)' \nonumber\\
& &{}\qquad\qquad\qquad - \frac{3C_2(R)}{4\pi}\frac{x\Sigma(x)}{x+\Sigma^2(x)}
\left(\frac{\alpha_s(x)}{x}\right)'^2=0,
\nonumber\\
\end{eqnarray}
with boundary conditions
\begin{eqnarray}
\Sigma'(0) & = &{} - \frac{3C_2(R)\alpha_s(0)}{8\pi\Sigma(0)}, \nonumber\\
\Sigma(\Lambda') & = & \frac{3C_2(R)\alpha_s(\Lambda')}{4\pi\Lambda'}
\int^{\Lambda'}_0dx\frac{x\Sigma(x)}{x+\Sigma^2(x)},
\end{eqnarray}
where $\alpha_s$ is the running coupling constant of QCD. $\Lambda'$ is an ultraviolet cutoff regularizing the integral, which should be taken $\Lambda^\prime \rightarrow \infty$ eventually.
	
Since the low energy behavior of the QCD running coupling constant is unclear to us, as a comparison, we take two models for $\alpha_s$. One is a segmented form of the coupling constant~\cite{Yang:2002hea}, which has a finite infrared limit (Model A):
\begin{eqnarray}
\alpha_s(p^2) & = & \frac{12\pi}{33-2N_f} \nonumber\\
& &{} \times\begin{cases}
a &  \ln  \frac{p^2}{\Lambda^2_{\rm QCD}}\leq b \\
a-c(2+  \ln  \frac{p^2}{\Lambda^2_{\rm QCD}})^2   &   b \leq  \ln  \frac{p^2}{\Lambda^2_{\rm QCD}}\leq d \\
\frac{1}{\ln (p^2/\Lambda^2_{\rm QCD})}  & d \leq \ln  \frac{p^2}{\Lambda^2_{\rm QCD}}
\end{cases}
,\nonumber
\end{eqnarray}
where $N_f=2$ and $\Lambda_{\rm QCD}=0.25$~GeV; $a,~b$ and $d$ are parameters of the model, and $c$ is an independent parameter keeping the strong coupling constant continuous at the boundary via the relation $c=(a-1/d)/(2 + d)^2$. We have found that the values of $d=4$ and $b=3$ give reasonable sets of results, so we fix them and focus on the variation of the results on the parameter $a$ alone.

\begin{figure}[h]
	\centering	
	\includegraphics[height=5.0cm]{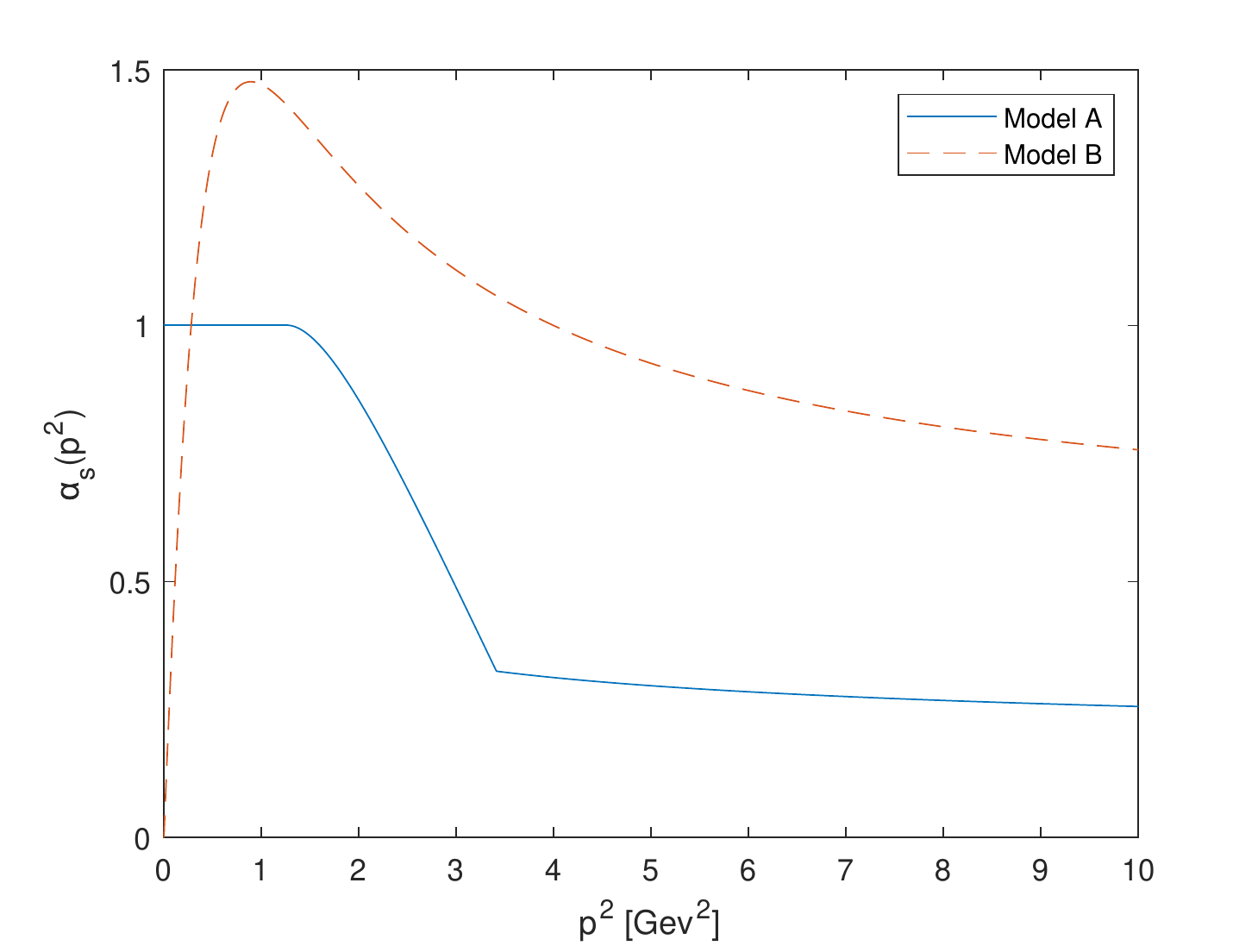}
	\caption{Running coupling constant of model A with $a=0.79$ and of model B with $a_0=0.58$. }	
	\label{rcc}
\end{figure}

The other form of the strong coupling constant is taken from the refined Gribov-Zwanziger formalism~\cite{Dudal:2012zx}, which has a vanishing infrared limit (Model B):
\begin{equation}
\alpha_s(p^2)=\alpha_0p^2\frac{p^2+M^2}{p^4+(M^2+m^2)p^2+\lambda^4},
\end{equation}
where $M^2=4.303~{\rm GeV}^2 , (M^2+m^2)= 0.526~{\rm GeV}^2 $ and $\lambda^4=0.4929~{\rm GeV}^4, \alpha_0$ is the parameter of the model. It should be noted that, different from Model A, $\alpha_s$ in this model does not respect the UV behavior of QCD. However, since the LECs are mostly controlled by the low energy behavior of QCD, this should not be a problem. Model A and B are sketched in Fig.~\ref{rcc}.

By solving the DSE, and choosing the parameters, we obtain $\Sigma(-p^2)$ for both models. Then LECs are calculated according to Eqs.~\eqref{eq:mHgH}-\eqref{eq:gHG}. It is clear that the integrals of the LECs have a physical ultraviolet cutoff $\Lambda_c$ which should be of the order of the chiral symmetry breaking scale. We have found that $\Lambda_c = 0.9$~GeV gives the best global-fitting results for the LECs. The results of LECs as well as the quark condensate $\langle\bar{q} q\rangle$ calculated with different parameters are shown in Tables~\ref{tab1} and \ref{tab2}. As expected, we find that the mass splitting between the chiral partners increases as the quark condensate increases. Our results directly show how the dynamics of the fundamental theory affects the LECs of the effective theory.

\begin{table*}[!tbp]
	\centering
	\caption{The heavy-light meson masses and the coupling constants calculated from Model A. $\Delta m=m_G-m_H$. ($b=3$,~$ d=4 $,~$\Lambda_{\rm QCD}=0.25$~GeV).}
	\setlength{\tabcolsep}{3mm}
	\begin{tabular}{ccccccccccc}
		\hline\hline
		$a$  &  ${}-\langle\bar{\psi}\psi\rangle$ (GeV$^{3}$) & $\Sigma(0)$~(GeV) & $f_{\pi}$(GeV) & $g_G $  & $g_H$ & $g_{HG}$ & $m_G$~(GeV) & $m_H$~(GeV) & $\Delta m $~(GeV) \\
		\hline
		0.81  & $(0.290)^3$ & 0.380 & 0.095 & 1.269 & 0.272 & 0.667 & 1.064 & 0.577 & 0.486 \\
		0.80  & $(0.285)^3$ & 0.357 & 0.092 & 1.194 & 0.249 & 0.694 & 1.015 & 0.564 & 0.451 \\
		0.79  & $(0.280)^3$ & 0.340 & 0.089 & 1.140 & 0.231 & 0.713 & 0.979 & 0.555 & 0.424 \\
		0.78  & $(0.276)^3$  & 0.323 & 0.086 & 1.089 & 0.214 & 0.730 & 0.947 & 0.546 & 0.400 \\
		0.77  & $(0.270)^3$  & 0.304 & 0.083 & 1.031 & 0.193 & 0.750 & 0.910 & 0.536 & 0.373 \\		
		\hline	\hline
	\end{tabular}%
	\label{tab1}%
\end{table*}%

\begin{table*}[!tbp]
	\centering
	\caption{The heavy-light meson masses and the coupling constants calculated from Model B. $\Delta m=m_G-m_H$. ($M^2=4.303~{\rm GeV}^2$, $M^2+m^2=0.526~{\rm GeV}^2$, $\lambda^4=0.4929~{\rm GeV}^2$).}
		\setlength{\tabcolsep}{3mm}
	\begin{tabular}{ccccccccccc}
		\hline\hline
			$a_0$  & ${}-\langle\bar{\psi}\psi\rangle$ (GeV$^{3}$) & $\Sigma(0)$~(GeV) & $f_{\pi}$(GeV) & $g_G $  & $g_H$ & $g_{HG}$ & $m_G$~(GeV) & $m_H$~(GeV) & $\Delta m $~(GeV) \\
			\hline
		0.60  & $(0.319)^3$  & 0.484 & 0.113 & 1.636 & 0.459 & 0.321 & 1.259 & 0.705 & 0.554 \\
		0.59  & $(0.316)^3$  & 0.450 & 0.111 & 1.526 & 0.434 & 0.378 & 1.177 & 0.684 & 0.494 \\
		0.58  & $(0.313)^3$  & 0.418 & 0.109 & 1.425 & 0.408 & 0.432 & 1.103 & 0.663 & 0.440 \\
		0.57  & $(0.309)^3$  & 0.386 & 0.106 & 1.330 & 0.381 & 0.482 & 1.035 & 0.643 & 0.392 \\
		0.56  & $(0.304)^3$  & 0.354 & 0.103 & 1.240 & 0.353 & 0.530 & 0.972 & 0.623 & 0.349 \\		
			\hline	\hline
	\end{tabular}%
	\label{tab2}%
\end{table*}%

For the mass splitting, we find that both Model A (with parameter $a=0.79$) and Model B (with parameter $a_0=0.58$) can give the results which agree with empirical data ~(\ref{eq:splitting}). To compare the masses of $H$ and $G$ doublets calculated here with experimental data of $D$ mesons, the charm quark mass which has been rotated away should restored. By using $m_c\approx 1.27$ GeV~\cite{ZylaPDG} we obtain $\tilde{m}_H\approx 1.825$ GeV and $\tilde{m}_G\approx 2.249$ GeV when using the data with parameter $a=0.79$ in Table \ref{tab1} or $\tilde{m}_H\approx 1.933$ GeV and $\tilde{m}_G\approx 2.373$ GeV when using the data with parameter $a=0.58$ in Table \ref{tab2}. These results are consistent with experimental data of spin-averaged masses of the charmed $D$ mesons~\eqref{eq:EXmass}. For the coupling constants, we find Model B with parameter $a_0=0.58$ is more favorable. In this case, the coupling $g_H=0.408$, which directly determines the width of the $D^{\ast} \to D \pi $ decay. By using the expression
\begin{eqnarray}
\Gamma(D^{\ast +}\to D\pi)=\frac{3}{2}\frac{g^2_{H} |P_{\pi}|^3}{12\pi f_\pi^2}
\end{eqnarray}
with $P_{\pi}$ being the three momentum of the decay products, we obtain the decay width $\Gamma(D^{\ast +}\to D\pi)=48$~KeV which is roughly comparable to the experimental result $  83.4 \pm 1.8$~KeV~\cite{ZylaPDG}.

In addition to the above intramultiplet transition, one can also calculate the intermultiplet transitions. The coupling constant responsible for these transitions is calculated as $g_{HG}=0.432$. From this value, the intermultiplet one-pion transition $ D_0^{\ast 0 +}\to D\pi$ is obtained as
\begin{eqnarray}
\Gamma(D_0^{\ast 0}\to D\pi)=\frac{3}{2}\frac{g^2_{HG} }{4\pi f_\pi^2} \frac{m_D}{m_{D^*_0}}|P_{\pi}|^3=183~\text{MeV},
\end{eqnarray}
which is at the same order as the experimental value $274\pm 40$~MeV~\cite{ZylaPDG}.

\begin{figure}[!t]
	\centering	
\includegraphics[height=6.5cm]{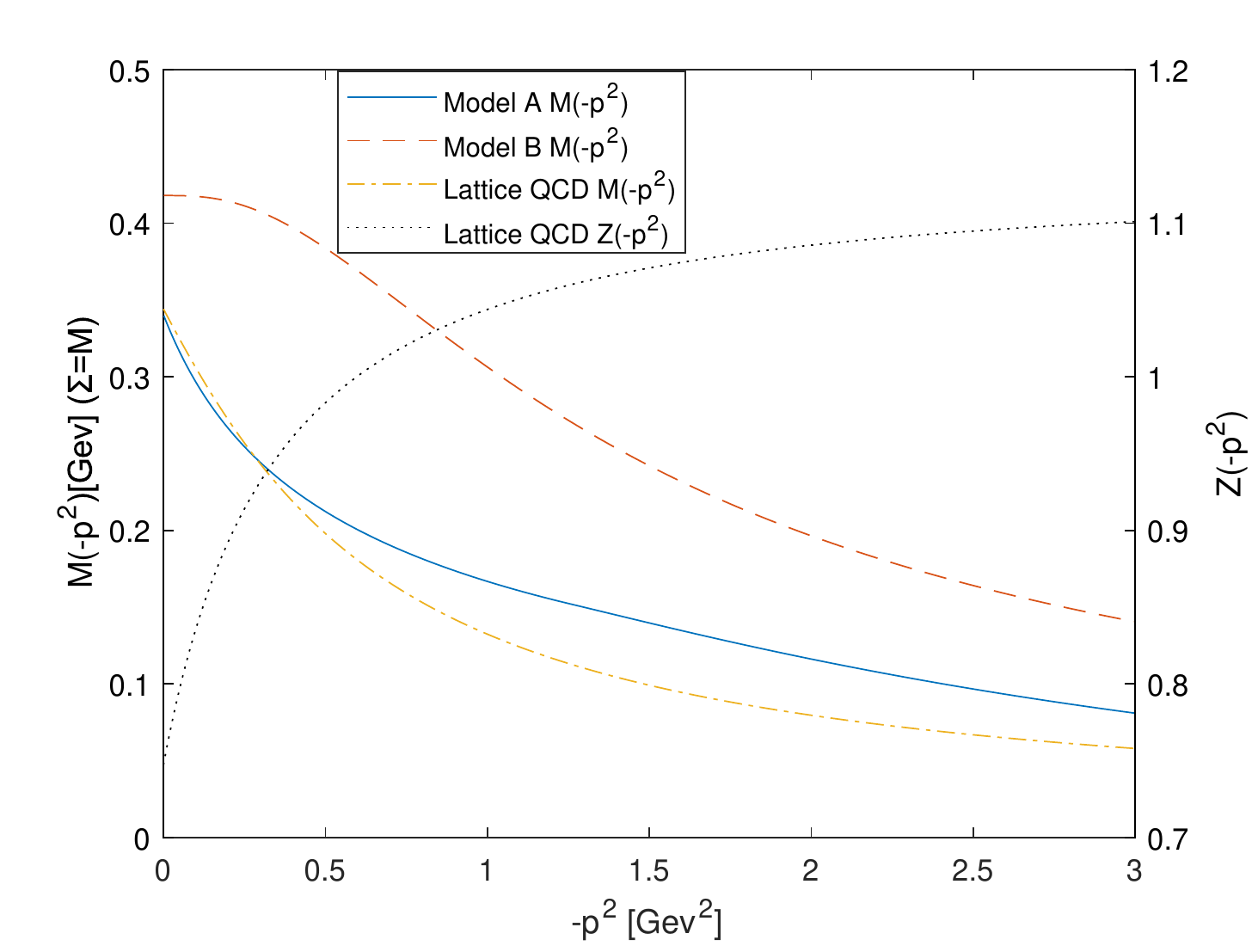}
\caption{The lattice fittings of $Z(-p^2)$ and $M(-p^2)$ given in Ref. \cite{Oliveira:2018ukh} and $\Sigma(-p^2)$ ($=M(-p^2)$) from DSE with Model A and Model B. }	
\label{selfenergy}
\end{figure}

Next, we consider the information from lattice QCD. In Ref.~\cite{Oliveira:2018ukh}, the authors fitted the lattice results (corresponding to $M_\pi=295$ MeV) for the quark wave function renormalization $Z(-p^2)$ and the running quark mass $M(-p^2)$. These functions are plotted in Fig.~\ref{selfenergy}. We shall use these fitted functions to calculate the LECs. The LECs including the contributions from $Z(-p^2)$ are given in Appendix B. $Z(-p^2)$ gives additional correction to the LECs, but, it also introduces additional uncertainties due to its dependence on the renormalization point (the running quark mass $M(-q^2)$ is renormalization independent and has no such uncertainties). As a comparison, we calculate LECs in two manners---one using the fitted $Z(-p^2)$ and the other using $Z(-p^2)=1$---in this lattice based analysis. The results are shown in Table~\ref{tab3}.

\begin{table*}[!tbp]
	\centering
	\caption{LECs calculated from lattice fittings given in Ref \cite{Oliveira:2018ukh}.}
	\setlength{\tabcolsep}{3mm}
	\begin{tabular}{cccccccc}
		\hline\hline
		 &   $f_{\pi}$ (GeV)& $g_G $  & $g_H$ & $g_{HG}$ & $m_G$~(GeV) & $m_H$~(GeV) & $\Delta m $~(GeV) \\
		\hline
		fitted $Z(-p^2)$   & 0.089 & 0.919 & 0.176 & 0.706 & 0.883 & 0.583 & 0.299 \\
		$Z(-p^2)=1$   & 0.089 & 1.267 & 0.215 & 0.757 & 1.125 & 0.537 & 0.588 \\		
		\hline	\hline
	\end{tabular}%
	\label{tab3}%
\end{table*}%

From Table~\ref{tab3} we find that the wave function renormalization affects the positive parity states more significantly. In turn, it affects the mass splitting $\Delta m$. The $\Delta m$'s given here deviate from the empirical value by about $140$ MeV. However, recalling that the mass splitting of $0^-$--$1^-$ and that of $0^+$--$1^+$ are both $\gtrsim 100$ MeV, these results are not strange. The real masses (i.e., with $m_c$ added up) are  $\tilde{m}_H\approx 1.853$ GeV and $\tilde{m}_G\approx 2.153$ GeV for the ``fitted $Z(-p^2)$" case and $\tilde{m}_H\approx 1.807$ GeV and $\tilde{m}_G\approx 2.395$ GeV for the ``$Z(-p^2)=1$" case. For the coupling constants, the results with lattice fittings are closer to those of model A than those of model B in the DSE based analysis.

In general, the lattice-based results are not so good as the best-fitted DSE results. But one should notice that lattice data themselves are still suffering from some uncertainties. For instance, the lattice fitted $Z(-p^2)$ and $M(-p^2)$ used here correspond to an overestimated pion mass $m_\pi=295$ MeV. So that the difference is acceptable.

\section{discussion and perspective}

\label{sec:dis}

In this paper, we derive the chiral effective Lagrangian of heavy-light mesons from QCD. The relationship between the LECs and the quark self-energy is obtained. Using the quark self-energy calculated from DSEs as well as fitted from lattice QCD, we calculated the values of the LECs in the Lagrangian. With properly chosen parameters, the numerical results are roughly consistent with the experimental data.

The deviations between our numerical results and the available empirical data can be understood as follows: First, as the leading order calculation in heavy quark expansion and large $N_c$ expansion, the results suffer from high order corrections from both $1/N_c$ terms and $1/M$ terms as we mentioned before. Second, the non-Tr ln term in the action has been omitted in our calculation for simplicity. However these terms generate contributions with complicated forms which in general cannot be described by the self-energy alone.

The scheme developed here for deriving the chiral effective Lagrangian of heavy-light mesons and calculating the LECs from QCD can be straightforwardly extended to include the $1/M$ corrections, higher order derivatives as well as excited heavy-light mesons. This is beyond the scope of the present work and will be reported elsewhere. In addition, it might be possible to extend the present work to the exotic heavy hadrons such as the tetraquark states. Such kind of study is in progress.

The last but not the least point we want to mention is that, the results shown here illustrate the quark condensate dependence of the LECs in the effective theory. When the change of the quark condensate due to the environment is known, one can easily translate our results to the change of the LECs therefore obtain the intrinsic environment, such as the density, dependence of the effective theory.

\appendix

\renewcommand{\appendixname}{Appendix}

\section{DERIVATION OF THE ACTION}

\label{sec:AppA}

In this Appendix, we derive Eq.~(\ref{action2}) from Eq.~(\ref{generating0}) in detail. Essentially, the procedure is summarized as integrating in the (pseudo-)Nambu-Goldstone boson fields, the heavy-light meson fields and integrating out the quark fields.

\subsection{Integrate in the (pseudo-)Nambu-Goldstone boson fields}

First, we focus on the (pseudo-)Nambu-Goldstone boson part in the effective Lagrangian. In the chiral limit considered in this work the QCD generating functional Eq. (\ref{generating0}) can be written as
\begin{widetext}
\begin{eqnarray}
Z[J] & = &\int \mathcal{D}q\mathcal{D} \bar q \mathcal{D}Q \mathcal{D} \bar Q  \exp \left\{ i\int d^4x \left[ \bar q (i\slashed{\partial}+J)q+ \bar Q (i\slashed{\partial}-M)Q \right]\right\}  \notag \\
& &{}~~\times\int \mathcal{D} G_{\mu}\varDelta_F(G_{\mu})\exp\left\{i \int d^4x\left[{}-\frac{1}{4}G_{\mu\nu}^iG^{\mu\nu}_i-\dfrac{1}{2\xi}[F^i(G_{\mu})]^2-g \mathcal{I}^{\mu}_i G^i_{\mu} \right] \right\},
\label{generating}
\end{eqnarray}
\end{widetext}
where $M$ is the mass matrix for the heavy quark fields. Here and in the following, we use $i, j, \cdots$ to represent color indices in the adjoint representation, and
$\mathcal{I}_i^{\mu}=\bar q \frac{\lambda_i}{2}\gamma^{\mu} q+\bar Q  \frac{\lambda_i}{2}\gamma^{\mu} Q$ to denote the quark composite operator. The external source $J(x)$ can be generally decomposed into scalar, pseudoscalar, vector, and axialvector
parts
\begin{eqnarray}
J(x) & = & {} -s(x)+ip(x)\gamma_5+\slashed v(x)+\slashed a(x)\gamma_5,
\end{eqnarray}
where $s(x)$, $p(x)$, $v_{\mu}(x)$, and $a_{\mu}(x) $ are Hermitian matrices in the flavor space, and the light quark masses have been absorbed into the definition of $s(x)$. The vector and axial-vector sources $v_{\mu}(x)$ and $a_{\mu}(x) $ are taken to be traceless. To introduce the pseudoscalar meson field $U=\Omega^2=e^{i\pi(x)/f_\pi}$ into the theory, following Ref.~\cite{Wang:1999cp}, we sandwich the following constant
\begin{eqnarray}
I & = &\int \mathcal{D}U \delta(U^{\dagger}U-1)\delta(\det U-1)\nonumber\\
& &{} \qquad\times \mathcal{F}[\mathcal{O}]\delta\left(\Omega\mathcal{O}^{\dagger}\Omega-\Omega^{\dagger}\mathcal{O}\Omega^{\dagger}\right),  \label{con1}
\end{eqnarray}
into the QCD generating functional~\eqref{generating}. In Eq.~\eqref{con1}, $\mathcal{F}[\mathcal{O}]\equiv\{\prod\limits_x\det\mathcal{O}\int\mathcal{D}\sigma\delta(\mathcal{O}^{\dagger}\mathcal{O}-\sigma^{\dagger}\sigma)\delta(\sigma-\sigma^{\dagger})\}^{-1}$, and $\mathcal{O}(x)=e^{-i[\theta(x)/N_f]}\mathrm{tr}_l[P_R B^T(x,x)]$. $B$ is the abbreviation of the bilocal composite light quark fields: $B^{ab}_{\eta\xi}(x,y)\equiv\frac{1}{N_c}\bar{q}_{\eta\alpha}^{a}(x)q_{\xi\alpha}^{b}(y)$, where $\eta, \xi, \cdots$ represent spinor indices, $a, b, \cdots$ denote flavor indices and $\alpha, \beta, \cdots$ stand for color indices in the fundamental representation. $N_f$ is the number of light flavors, and $\mathrm{tr}_l$ denotes tracing over the spinor space. $\theta(x)$ is defined as $e^{i2\theta(x)}\equiv \frac{\det \mathrm{tr}_l[P_R B^T(x,x)]}{\det \mathrm{tr}_l[P_L B^T(x,x)]}$. Inserting Eq.~(\ref{con1}) into Eq.~(\ref{generating}) we get
\begin{widetext}
\begin{eqnarray}
\label{generating4}
	Z[J] & = & \int \mathcal{D}q\mathcal{D} \bar q \mathcal{D} Q \mathcal{D} \bar Q \mathcal{D} G_{\mu}\varDelta_F(G_{\mu})\mathcal{D}U \delta(U^{\dagger}U-1)\delta(\det U-1)\delta(\Omega\mathcal{O}^{\dagger}\Omega-\Omega^{\dagger}\mathcal{O}\Omega^{\dagger})  \notag \\
& & \qquad \times \exp\biggl\{ i\int d^4 x\biggl [ \bar q (i\slashed{\partial}+J)q+\bar  Q (i\slashed{\partial}-M) Q -\frac{1}{4}G_{\mu\nu}^iG^{\mu\nu}_i-\dfrac{1}{2\xi}[F^i(G_{\mu})]^2-g \mathcal{I}^{\mu}_i G^i_{\mu}\biggr] +i\Gamma_I[B]\biggr\},
\end{eqnarray}
\end{widetext}
where $e^{i\Gamma_I[B]}\equiv \mathcal{F}[\mathcal{O}]$.

The exponential part in Eq.~(\ref{generating4}) is invariant under the chiral rotation
\begin{eqnarray}
q & \rightarrow & q_{\Omega}=(\Omega P_L+\Omega^{\dagger}P_R) q, \nonumber\\
\bar q & \rightarrow & \bar q_{\Omega}=\bar q(\Omega P_L+\Omega^{\dagger}P_R),
\end{eqnarray}
if the external field $J(x)$ undergoes the following transformation
\begin{eqnarray}
J(x) & \to& J_{\Omega}(x)=[\Omega(x)P_{R}+\Omega^{\dagger}(x)P_L]\nonumber\\
& &{} \qquad\qquad \times [J(x)+i\slashed \partial][\Omega(x)P_{R}+\Omega^{\dagger}(x)P_L].
\nonumber
\end{eqnarray}	
In addition, one can find
\be
\Omega \mathcal{O}^{\dagger}\Omega-\Omega ^{\dagger}\mathcal{O} \Omega^{\dagger}=\mathcal{O}^{\dagger}_{\Omega}-\mathcal{O}_{\Omega}.\nonumber
\ee
		
In this work, we concentrate on the normal part of the chiral Lagrangian of the heavy-light mesons and ignore the chiral anomaly. In this case, the generating functional can be reexpressed in terms of rotated fields as: 	
\begin{widetext}	
\begin{eqnarray}
Z[J] & = & \int \mathcal{D} q_{\Omega}\mathcal{D} \bar  q_{\Omega} \mathcal{D} Q \mathcal{D} \bar Q \mathcal{D} G_{\mu}\varDelta_F(G_{\mu})\mathcal{D}\bar U \delta(\mathcal{O}_{\Omega}^{\dagger}-\mathcal{O}_{\Omega})\notag \\
& &{} ~~~~ \exp \biggl\{ i\int d^4 x\biggl [\bar  q_{\Omega} (i\slashed{\partial}+J_{\Omega}) q_{\Omega}+ \bar  Q (i\slashed{\partial}-M) Q -\frac{1}{4}G_{\mu\nu}^iG^{\mu\nu}_i-\dfrac{1}{2\xi}[F^i(G_{\mu})]^2-g \mathcal{I}_{i\Omega}^{\mu} G^i_{\mu}\biggr ]+i\Gamma_I[B_{\Omega} ]\biggr\},\label{generating3}
\end{eqnarray}
\end{widetext}
where $ \mathcal{D}\bar U =\mathcal{D}U \delta(U^{\dagger}U-1)\delta(\det U-1) $.

Now we integrate out the gluon fields, which generates pure Yang-Mills gluon Green's functions $G^{i_1 \cdots i_n}_{\mu_1 \cdots \mu_n}(x_1, \cdots,x_n )$ in the action. By Fierz reordering, we can further diagonalize the color indices of the quark fields, and get
\begin{widetext}
\begin{eqnarray}
\label{congpai}
& & G^{i_1 \cdots i_n}_{\mu_1 \cdots \mu_n}(x_1, \cdots,x_n )\left[\bar{ \psi} ^{a_1}_{\alpha_1}\left(\frac{\lambda_{i_1}}{2}\right)_{\alpha_1\beta_1}\gamma^{\mu_1} \psi^{a_1}_{\beta_1}(x_1)\right] \cdots \left[\bar { \psi}^{a_n}_{\alpha_n}\left(\frac{\lambda_{i_n}}{2}\right)_{\alpha_n\beta_n}\gamma^{\mu_n} \psi^{a_n}_{\beta_n}(x_n)\right]  \notag\\
&& \qquad\qquad\qquad =\int d^4x'_1 \cdots d^4 x'_n g^{n-2} \bar G^{\sigma_1 \cdots \sigma_n}_{\rho_1 \cdots \rho_n}(x_1,x'_1, \cdots ,x_n,x'_n)\times \bar { \psi}^{\sigma_1}_{\alpha_1}(x_1)  \psi^{\rho_1}_{\alpha_1}(x'_1)\cdots \bar { \psi}^{\sigma_n}_{\alpha_n}(x_n)  \psi^{\rho_n}_{\alpha_n}(x'_n), 		
\end{eqnarray}
where $\psi$ stands for both the light and heavy quarks, i.e., $\psi=(q,Q)$. $\bar G^{\sigma_1 \cdots \sigma_n}_{\rho_1 \cdots \rho_n}(x_1,x'_1, \cdots ,x_n,x'_n)$ is a generalized gluon Green's function. Then, the generating functional becomes
\begin{eqnarray}
Z[J] & = & \int \mathcal{D} q \mathcal{D} \bar q \mathcal{D}  Q \mathcal{D}\bar  Q \mathcal{D}\bar U \delta(\mathcal{O}^{\dagger}-\mathcal{O}) \nonumber\\
& &{} \times \exp\biggl\{i\int d^4x \biggl[ \bar  q (i\slashed{\partial}+J_\Omega) q + \bar  Q (i\slashed{\partial}-M) Q \biggr]+i\Gamma_I[B] \notag \\
& &\qquad\qquad{} +\sum\limits_{n=2}^{\infty}\int d^4x_1 \cdots d^4x'_n d^4x'_1 \cdots d^4 x'_n \dfrac{(-i)^n (N_cg^2)^{n-1}}{n!} \bar G^{\sigma_1 \cdots \sigma_n}_{\rho_1 \cdots \rho_n}(x_1,x'_1, \cdots ,x_n,x'_n) \notag \\
& &\qquad\qquad\qquad\qquad{}\times \bar {\psi}^{\sigma_1}_{\alpha_1}(x_1)   \psi^{\rho_1}_{\alpha_1}(x'_1)\cdots \bar {\psi}^{\sigma_n}_{\alpha_n}(x_n) \psi^{\rho_n}_{\alpha_n}(x'_n) \biggl\}.
\label{generating2}
\end{eqnarray}
\end{widetext}

\subsection{Integrate in the heavy-light meson fields}

Next, we introduce the heavy-light meson fields into the system by considering the heavy quark symmetry. Following the standard heavy quark effective theory (HQET) ( see,.e.g., Ref.~\cite{Manohar:2000dt} for a pedagogical discussion), we introduce the velocity-dependent heavy quark field by using the following substitution
\begin{eqnarray}
\label{Qprime}
 Q'(x)&\equiv e^{iMv\cdot x} Q(x),
\end{eqnarray}
where $v^{\mu}$ is related to the heavy quark momentum $p^\mu$ by $p^\mu=M v^\mu+k^\mu$ with $k^\mu$ the residue momentum. Introducing the projection operator $(1\pm \slashed{v})/2$, one can decompose the heavy quark field as
\begin{eqnarray}
N_v(x) & = & \frac{1+\slashed{v}}{2}e^{iMv\cdot x} Q(x), \nonumber\\
\mathcal{N}_v(x) & = & \frac{1-\slashed{v}}{2}e^{iMv\cdot x} Q(x).
\end{eqnarray}
The $N_v$ field is the large component of the quark field which survives in the heavy quark limit whereas the $\mathcal{N}_v$ field is the small component of the quark field which disappears in the heavy quark limit.
Applying Eq.~(\ref{Qprime}) to Eq.~(\ref{generating2}) amounts to the replacement $\bar  Q (i\slashed{\partial}-M) Q\rightarrow\bar  Q' (i\slashed{\partial}+M\slashed v-M)Q'$.

To proceed, we introduce a bilocal auxiliary field $ \Phi^{ab}_{\eta \xi }$ by sandwiching the following constant into Eq.~\eqref{generating2}:
\begin{equation}
\int \mathcal{D}\Phi\delta(N_c\Phi^{ab}_{\eta\xi}(x,x')-\bar{ \psi}'^{a}_{\eta}(x)  \psi'^{b}_{\xi}(x')),
\end{equation}
where $\psi'=(q, N_v)$ in the sense that only the contribution from $N_v$---the large component of the heavy quark field---is considered. ~\footnote{Here, we focus on the physics in the heavy quark limit. The finite corrections to the LECs calculated later and the form of the effective theory from the heavy quark mass can be easily included in our approach along Ref.~\cite{Wu:2005fd}.}
The generating functional then becomes
\begin{widetext}
\begin{eqnarray}
Z[J] & = & \int \mathcal{D} q\mathcal{D} \bar q \mathcal{D}  Q' \mathcal{D}\bar  Q'\mathcal{D}\bar U\mathcal{D}\Phi \delta(\mathcal{O}^{\dagger}-\mathcal{O})\delta(N_c\Phi^{(a\eta)(b\xi)}(x,x')-\bar{\psi}'^{a}_{\eta}(x) \psi'^{b}_{\xi}(x') ) \notag \\
& &{}\times\exp i\biggl\{\int d^4 x \biggl[ \bar  q (i\slashed{\partial}+J_\Omega) q+ \bar  Q' (i\slashed{\partial}+M\slashed v-M)Q' \biggr] +i\Gamma_I[\Phi]+N_c\sum\limits_{n=2}^{\infty}\int d^4x_1 \cdots d^4x'_n d^4x'_1 \cdots d^4 x'_n\notag \\
& & {}\qquad\qquad  \times\dfrac{(-i)^n (N_cg^2)^{n-1}}{n!} \bar G^{\sigma_1 \cdots \sigma_n}_{\rho_1 \cdots \rho_n}(x_1,x'_1, \cdots ,x_n,x'_n)\times \Phi^{\sigma_1\rho_1}(x_1,x'_1) \cdots \Phi^{\sigma_n\rho_n}(x_n,x_n')\biggr\},
\end{eqnarray}
where we have replaced the bilocal quark fields with $\Phi$ in the spirit of the heavy quark limit.

The $\delta$ function  can be further expressed in the
Fourier representation
\begin{eqnarray}
	\delta{(N_c\Phi^{ab}_{\eta\xi}(x,x')-\bar{\psi}'^{a}_{\eta}(x) \psi'^{b}_{\xi}(x') )}\sim \int \mathcal{D}\Pi \exp^{i\int d^4 xd^4 x'\Pi^{ab}_{\eta\xi}(x,x')[N_c\Phi^{ab}_{\eta\xi}(x,x')-\bar{\psi}'^{a}_{\eta}(x) \psi'^{b}_{\xi}(x')]}.
	\label{delta}
\end{eqnarray}
Then the generating functional becomes
\begin{eqnarray}
\label{generating5}
 Z[J] & = & \int \mathcal{D} q\mathcal{D} \bar q\mathcal{D}  Q' \mathcal{D}\bar  Q'\mathcal{D}\bar U\mathcal{D}\Phi\mathcal{D}\Pi  \delta(\mathcal{O}^{\dagger}-\mathcal{O}) \notag\\
 & & {} \times\exp\biggl\{i\int d^4x \biggl[ \bar  q (i\slashed{\partial}+J_\Omega) q+ \bar  Q' (i\slashed{\partial}+M\slashed v-M)Q'\biggr] +i\Gamma_I[\Phi]+iN_c\bar G(\Phi)\notag\\
 & & {} \qquad\qquad  +i\int d^4 xd^4 x'\biggl[N_c\Pi^{ab}_{\eta\xi}(x,x')\Phi^{ab}_{\eta\xi}(x,x')-\bar{\psi}'(x)\Pi(x,x') \psi'(x')\biggr] \biggr\},
\end{eqnarray}
where
\begin{eqnarray}
\bar G(\Phi)=\sum\limits_{n=2}^{\infty}\int d^4x_1 \cdots d^4x'_n d^4x'_1 \cdots d^4 x'_n \dfrac{(-i)^n (N_cg^2)^{n-1}}{n!} \bar G^{\sigma_1 \cdots \sigma_n}_{\rho_1 \cdots \rho_n}(x_1,x'_1, \cdots ,x_n,x'_n) \Phi^{\sigma_1\rho_1}(x_1,x'_1) \cdots \Phi^{\sigma_n\rho_n}(x_n,x_n').
\nonumber
\end{eqnarray}
\end{widetext}

Since $N_v(x)=\frac{1+\slashed{v}}{2}Q'(x)$, only the ``positive projected" part (the part projected by $\frac{1+\slashed{v}}{2}$) of $\Pi^{ab}(x,y)$ contributes. For example, for $\Pi^{qQ}$,
\begin{eqnarray}
\bar{q}(x)\Pi^{qQ}(x,y)N_v(y) & = & \bar{q}(x)\Pi^{qQ}(x,y)\frac{1+\slashed{v}}{2}Q'(y)\nonumber\\
& = & \bar{q}(x)\Pi^{qQ}_{+}(x,y)Q'(y),
\end{eqnarray}
where $\Pi^{qQ}_{\pm} = \Pi^{qQ}\frac{1\pm\slashed{v}}{2}$. Since we are only interested in the leading contribution from heavy quark expansion, $\Pi^{qQ}_{-}$ has no contributions here.
So eventually, we can keep only the positive projected parts of the $\Phi$ and $\Pi$ fields for their heavy flavor components in the generating functional.


\subsection{Integrate out quark fields}

Now, we are ready to integrate out quark fields $q$ and $N_{v}$. The standard procedure yields
\begin{widetext}
\begin{eqnarray}
Z[J] & = & \int\mathcal{D}\bar U\mathcal{D}\Phi\mathcal{D}\Pi  \delta(\mathcal{O}^{\dagger}-\mathcal{O}) \nonumber\\
& &{} \times\exp\biggl\{iN_c\biggl[-i \mathrm{Tr}^\prime \ln[ i\slashed{\partial} I_1+J'_\Omega +  (i\slashed{\partial}+M\slashed v-M)I_4 -\Pi ]\biggr] +iN_c \mathrm{Tr}^\prime(\Pi\Phi^T) +iN_c\bar G(\Phi) + i\Gamma_I[\Phi] \biggr\},
\end{eqnarray}
\end{widetext}
where we have defined the functional trace $\mathrm{Tr}^\prime$ taking over the flavor space, spinor space and coordinate space and $I_1$ and $I_4$ are the following matrices in the flavor space
\[I_1=\begin{pmatrix}
1 & 0 & 0\\
0 & 1 & 0\\
0 & 0 & 0
\end{pmatrix}, \quad I_4=\begin{pmatrix}
0 & 0 & 0\\
0 & 0 & 0\\
0& 0 & 1
\end{pmatrix}. \]
In this work we only consider two light flavors plus one heavy flavor. The generalization to include more flavors is straightforward. $J'_\Omega$ is the extension of $J_\Omega$ into the whole flavor space with $J'^{Qq}_\Omega=J'^{qQ}_\Omega=J'^{QQ}_\Omega=0$.

In flavor space the matrix $\Pi$ can be decomposed as
\begin{eqnarray}
\Pi=\Pi_1+\Pi_2+\Pi_3+\Pi_4,
\end{eqnarray}
where
\begin{eqnarray}
\Pi_1=\begin{pmatrix}
\Pi^{qq}_{2\times2}  & 0\\
0 & 0
\end{pmatrix}, \quad \Pi_2=\begin{pmatrix}
0  & 0\\
\Pi^{Qq}_{1\times2} & 0
\end{pmatrix},
\nonumber\\
\quad \Pi_3=\begin{pmatrix}
0  & \Pi^{qQ}_{2\times1}\\
0 & 0
\end{pmatrix},\quad \Pi_4=\begin{pmatrix}
0 & 0\\
0 & \Pi^{QQ}
\end{pmatrix}.
\end{eqnarray}
A similar decomposition holds for $\Phi$. Tracing back to Eq. (\ref{generating5}) and from the term $\bar{\psi}'(x)\Pi(x,x') \psi'(x')$, one can find that $\Pi$ directly couples to quark-anti-quark pairs and has the same transformation properties as the composite quark fields, so that it is reasonable to identify $\Pi_2$ and $\Pi_3$ as the bosonic interpolating fields for the heavy-light mesons. However, since $\Pi$ is a bilocal field, to get a local effective Lagrangian, we need a suitable localization on $\Pi_2$ and $\Pi_3$ fields. Here we take the following localization conditions which are essentially the point coupling between quarks~\cite{Tandy:1997qf}
\begin{eqnarray}
\Pi_2(x,y) & = & \Pi_2(x)\delta(x-y), \nonumber\\
\Pi_3(x,y) & = & \Pi_3(x)\delta(x-y).
\end{eqnarray}

It is easy to check that $\Phi$ and $\Pi$ have the following properties:
\be
& & \gamma^0[\Phi^{Tba}(y,x)]^{\dagger}\gamma^0 =\Phi^{Tab}(x,y),\nonumber\\
& & \gamma^0[\Pi^{ab}(x,y)]^{\dagger}\gamma^0=\Pi^{ba}(y,x).\nonumber
\ee
So that
$$\bar{\Pi}_2(x,x)\equiv\gamma^0\Pi_2^{\dagger }(x,x)\gamma^0=\Pi_{3}(x,x).$$
Then the generating functional can be written as
\begin{widetext}
\begin{eqnarray}
\label{gf}
Z[J] & = & \int\mathcal{D}\bar U\mathcal{D}\Phi\mathcal{D}\Pi  \exp\biggl\{i\Gamma_I[\Phi]\nonumber\\
& &{} \qquad\qquad\qquad\qquad  +iN_c\Big\{\mathrm{Tr}^\prime(\Pi\Phi^T) +\bar G(\Phi)+\int d^4x\mathrm{Tr}\left\{\Xi\left[\left(-i\sin\frac{\theta}{N_f}+\cos\frac{\theta}{N_f}\right)\Phi^T\right]\right\}\notag\\
& &{}\qquad\qquad\qquad\qquad\qquad\qquad -i \mathrm{Tr}^\prime \ln(i\slashed{\partial}I_1 +J'_\Omega
+ (i\slashed{\partial}+M\slashed v-M)I_4  -\Pi_1-\Pi_2-\bar{\Pi}_2 -\Pi_4)\Big\} \biggr\},
\end{eqnarray}
\end{widetext}
where the $\delta(\mathcal{O}^{\dagger}-\mathcal{O}) $ term has been reexpressed as
\begin{eqnarray}
\delta(\mathcal{O}^{\dagger}-\mathcal{O})=\int \mathcal{D}\Xi e^{iN_c\int d^4 x \mathrm{Tr}\{\Xi (x)[\Theta\Phi^T(x,x)]\}},
\end{eqnarray}
with $\Xi^{ab}(x)$ being a new auxiliary field and
\begin{eqnarray}
\Theta & \equiv & (-i\sin\theta(x)/N_f+\cos\theta(x)/N_f).
\end{eqnarray}

Now, by integrating out the fields $\Phi$, $\Xi$, $\Pi_1$ and $\Pi_4$, we can obtain the action, denoted as $S[U,\Pi_2,\bar{\Pi}_2]$, for the chiral effective theory with heavy-light mesons
\begin{eqnarray}
Z[J]=\int\mathcal{D}\bar U\mathcal{D}\Pi_2\mathcal{D}\bar{\Pi}_2 \exp\{ i S[U,\Pi_2,\bar{\Pi}_2] \},
\end{eqnarray}
where
\begin{widetext}
\begin{eqnarray}
e^{iS} & \equiv & \int\mathcal{D}\Phi\mathcal{D}\Pi_1\mathcal{D}\Pi_4\mathcal{D}\Xi \exp\biggl\{i\Gamma_I[\Phi]+iN_c\Big\{\mathrm{Tr}^\prime(\Pi\Phi^T) +\bar G(\Phi)+\int d^4x\mathrm{Tr}\{\Xi [(-i\sin\frac{\theta}{N_f}+\cos\frac{\theta}{N_f})\Phi^T]\}\notag\\
& &{}\qquad\qquad\qquad\qquad\qquad\qquad\qquad\qquad\quad -i \mathrm{Tr}^\prime \ln(i\slashed{\partial}I_1 +J'_\Omega
+ (i\slashed{\partial}+M\slashed v-M)I_4  -\Pi_1-\Pi_2-\bar{\Pi}_2 -\Pi_4)\Big\} \biggr\}.
\end{eqnarray}
\end{widetext}
The heavy-light meson effective theory can be derived by expanding the action $S$ with respect to the Goldstone boson field $U$ and the heavy-light meson fields $\Pi_2,\bar{\Pi}_2$.

\subsection{The action in the large $N_c$ Limit }

The action defined in the previous subsection is not practically useful. In order to calculate the coefficients in the chiral Lagrangian, we make a further approximation, namely, keeping the leading order of the large $N_c$ expansion. Under large $N_c$ limit, the effective action is simply the corresponding classical action, so we have
\begin{widetext}
\begin{eqnarray}
\label{action1}
S[U,\Pi_2,\bar{\Pi}_2] & = & N_c\Bigg\{\mathrm{Tr}^\prime(\Pi_{c}\Phi^T_c) +\bar G(\Phi_c)+\int d^4x\mathrm{Tr}\left\{\Xi_c \left[\left(-i\sin\frac{\theta(\Phi_c)}{N_f}+\cos\frac{\theta(\Phi_c)}{N_f}\right)\Phi^T_c\right]\right\}\notag\\
&& {} \qquad -i \mathrm{Tr}^\prime \ln\left[i\slashed{\partial}I_1 +J'_\Omega
+ \left(i\slashed{\partial}+M\slashed v-M\right)I_4  -\Pi_{1c}-\Pi_2-\bar{\Pi}_2 -\Pi_{4c}\right]\Bigg\},
\end{eqnarray}
\end{widetext}
where $\Phi_c, \Xi_c, \Pi_{1c}, \Pi_{4c}$ are classical fields satisfying the saddle point equations
\begin{equation}
\label{saddle}
\frac{\delta S}{\delta \Phi_c}=\frac{\delta S}{\delta \Xi_c}=\frac{\delta S}{\delta \Pi_{1c}}=\frac{\delta S}{\delta \Pi_{4c}}=0.
\end{equation}
and $\Gamma_I$ term has been ignored because it is of $O(1/N_c)$~\cite{Wang:1999cp}.

These saddle point equations provide important information. For instance, equations $\frac{\delta S}{\delta \Pi_{1c}}=0$ and $\frac{\delta S}{\delta \Phi_{1c}}=0$  generate the coupled equations
\begin{equation}
\label{DS1}
\Phi^{Tab}_{1c\eta\xi}(x,y)=-i\left[(i\slashed\partial+J'_\Omega-\Pi_{1c})^{-1}\right]^{ab}_{\eta\xi}(x,y),
\end{equation}
with
\begin{widetext}
\begin{eqnarray}
\Pi^{ab}_{1c\eta\xi}(x,y) & = &{} -\sum\limits_{n=1}^{\infty}\int d^4x_1 \cdots d^4x'_n d^4x'_1 \cdots d^4 x'_n \dfrac{(-i)^{n+1} (N_cg^2)^{n}}{n!}\notag\\
& &{} \qquad \qquad \times
\bar{G}_{\eta\eta_1\cdots\eta_n,\xi\xi_1\cdots\xi_n}^{aa_1\cdots a_n,bb_1\cdots b_n}(x,y,x_1,x'_1,\cdots,x_n,x'_n) \Phi^{a_1b_1}_{1c\eta_1\xi_1}(x_1,x'_1)\cdots \Phi^{a_nb_n}_{1c\eta_n\xi_n}(x_n,x'_n),\label{DS2}
\end{eqnarray}
\end{widetext}
where the term involving $\Xi$ has been omitted because it vanishes once the external sources $J$ are turned off~\cite{Wang:1999cp}. When $J_\Omega^\prime$ is turned off, the coupled equations (\ref{DS1}) and (\ref{DS2}) are nothing but the DSEs for the quark propagators with $\Pi_{1c}$ being the self-energy for light quarks. So that we rewrite $\Pi_{1c}(x,y)$ as $\bar{\Sigma}(x-y)I_1$.
Along the same procedure, the DSE for the heavy quark propagator which depends on the self-energy of heavy quark, $\Pi_{4c}$, can be obtained. However, since the contribution from the heavy quark self-energy is less significant than the light ones, we will simply ignore it.

The direct calculation of the LECs in the effective Lagrangian from action~(\ref{action1}) is not so easy, if not impossible, because the fields $\Phi_c$ and $\Xi_c$ are functionals of $U$, $\Pi_2$ and $\bar{\Pi}_2$ through the saddle point equations (\ref{saddle}). To proceed, we follow Ref.~\cite{Yang:2002hea} and keep only the ``$\mathrm{Tr}\ln$" term in the action in the spirit of the dynamical perturbation which works well in the calculation of pion decay constant~\cite{Pagels:1979hd}. Then the action becomes
\begin{eqnarray}
\label{action2APP}
S[U,\Pi_2,\bar{\Pi}_2] & = &
{} -i N_c \mathrm{Tr}^\prime \ln\left\{(i\slashed{\partial}-\bar{\Sigma})I_1 +J'_\Omega \right.\nonumber\\
& &\left. {} \qquad\qquad\qquad + (i\slashed{\partial}+M\slashed v-M)I_4\right.\nonumber\\
& &\left. {} \qquad\qquad\qquad  -\Pi_2-\bar{\Pi}_2 \right\}.
\end{eqnarray}
It should be noted that although the contribution from gluon fields is not explicit appeared in the action, it is included in the quark self-energy $\bar{\Sigma}$, which requires the application of the DSE.

\section{FORMULA FOR LECS WITH $Z(-p^2)$ EFFECT}

\label{sec:AppB}

The quark propagator can be generally written as
\begin{eqnarray}
S(p) & = &\frac{i}{A(p^2) \slashed  p\ -  B(-p^2) }
= i \, \frac{A(-p^2) \slashed  p\ + B(-p^2) }{ A^2(-p^2) \,  p^2 - B^2(-p^2)} \nonumber \\
& = &
i \, Z(-p^2) \, \frac{ \slashed  p\ + M(-p^2) }{ p^2 - M^2(-p^2)},\nonumber
\end{eqnarray}
where $Z(-p^2) = 1/A(-p^2)$ stands for the quark wave function renormalization and $M(-p^2) = B(-p^2) / A(-p^2)$ is the renormalization group invariant running quark mass.

Using this full expression of the quark propagator, we obtain the LECs as follows:
\begin{eqnarray}
m_H & = & \frac{iN_c}{Z_H}  \int\frac{d^4 p}{(2\pi)^4}  \biggr[{}\frac{1}{p^2-M^2} - \frac{M}{(p^2-M^2)v\cdot p}\biggr]Z,  \nonumber\\
g_{H} & = & -\frac{iN_c}{Z_H}\int\frac{d^4 p}{(2\pi)^4}\biggl[\frac{M^2+\frac{1}{3}p^2}{(p^2-M^2)^2 v\cdot p}  - \frac{2M}{(p^2-M^2)^2 } \biggr]Z^2,\nonumber\\
Z_H & = & iN_c\int\frac{d^4 p}{(2\pi)^4} \biggr[ \left(-\frac{1}{(p^2-M^2) v\cdot p} - \frac{2M}{(p^2-M^2)^2}\right ) Z \nonumber\\
&&\qquad\qquad\qquad-\frac{2Z M'(p^2+M^2)}{(p^2-M^2)^2} -\frac{2Z' M }{p^2-M^2} \biggr]\nonumber\\
m_G & = & \frac{iN_c}{Z_G}\int\frac{d^4 p}{(2\pi)^4} \biggr[\frac{1}{p^2-M^2} +\frac{M}{(p^2-M^2)v\cdot p}\biggr]Z, \nonumber\\
g_{G} & = & -\frac{iN_c}{Z_G}\int\frac{d^4 p}{(2\pi)^4}\biggl[\frac{M^2+\frac{1}{3}p^2}{(p^2-M^2)^2 v\cdot p}  + \frac{2M}{(p^2-M^2)^2 } \biggr]Z^2,\nonumber\\
Z_G & = &iN_c\int\frac{d^4 p}{(2\pi)^4} \biggr[ \left(-\frac{1}{(p^2-M^2) v\cdot p} + \frac{2M}{(p^2-M^2)^2}\right ) Z \nonumber\\
&&\qquad\qquad\qquad+\frac{2Z M'(p^2+M^2)}{(p^2-M^2)^2} +\frac{2Z' M }{p^2-M^2} \biggr],\nonumber\\
g_{H G} & = &  - i \frac{N_c}{\sqrt{Z_H Z_G}}\int\frac{d^4 p}{(2\pi)^4}\biggl[\frac{M^2+p^2}{(p^2-M^2)^2 v\cdot p} \biggr]Z^2.
\end{eqnarray}

\acknowledgments

The work of Y.~L. M. was supported in part by National Science Foundation of China (NSFC) under Grant No. 11875147 and No.11475071. Q. W. was supported by the National Science Foundation of China (NSFC) under Grant No. 11475092.

\bibliography{paper}

\end{document}